\documentclass[10pt,journal,compsoc]{IEEEtran}

\usepackage{balance}
\usepackage{booktabs} 
\usepackage{graphicx}
\usepackage{url}
\usepackage{color-edits}

\addauthor{pg}{red}
\addauthor{th}{blue}
\addauthor{ef}{green}
\addauthor{kl}{cyan}
\addauthor{nt}{purple}
\addauthor{ad}{orange}
\addauthor{aa}{red}
\addauthor{nb}{magenta}

\ifCLASSOPTIONcompsoc
  \usepackage[nocompress]{cite}
\else
  \usepackage{cite}
\fi

\usepackage{amsmath}
\usepackage{amsfonts}

\ifCLASSOPTIONcompsoc
\usepackage[caption=false,font=footnotesize,labelfont=sf,textfont=sf]{subfig}
\else
\usepackage[caption=false,font=footnotesize]{subfig}
\fi

\begin{document}
 \title{Discovering Signals from Web Sources to Predict Cyber Attacks}

\author{Palash Goyal*, 
			 KSM Tozammel Hossain*,
             Ashok Deb,
             Nazgol Tavabi,
             Nathan Bartley,
             Andr\'es Abeliuk,
             Emilio Ferrara
			 and Kristina Lerman 
\IEEEcompsocitemizethanks{\IEEEcompsocthanksitem Authors are with the Department of Computer Science, University of Southern California (USC), and with the USC Information Sciences Institute. *P. Goyal, T. Hossein equally contributed to this article.\protect\\
E-mail: goyal@isi.edu
}
\thanks{Manuscript received May 12, 2018; revised May 12, 2018.}}

\markboth{IEEE Systems,~Vol.~X, No.~X, August~XXXX}%
{Shell \MakeLowercase{\textit{et al.}}: Data Driven Discovery of Web Sources to Forecast Cyber Attacks}
\IEEEtitleabstractindextext{%
\begin{abstract}
Cyber attacks are growing in frequency and severity. Over the past year alone we have witnessed massive data breaches that stole personal information of millions of people and wide-scale ransomware attacks that paralyzed critical infrastructure of several countries. Combating the rising cyber threat calls for a multi-pronged strategy, which includes predicting when these attacks will occur. The intuition driving our approach is this: during the planning and preparation stages, hackers leave digital traces of their activities on both the surface web and dark web in the form of discussions on platforms like hacker forums, social media, blogs and the like. These data provide predictive signals that allow anticipating cyber attacks.  In this paper, we describe machine learning techniques based on deep neural networks and autoregressive time series models that leverage external signals from publicly available Web sources to forecast cyber attacks. Performance of our framework across ground truth data over real-world forecasting tasks shows that our methods yield a significant lift or increase of F1 for the top signals on predicted cyber attacks. Our results suggest that, when deployed, our system will be able to provide an effective line of defense against various types of targeted cyber attacks. 
\end{abstract}

}

\maketitle

\IEEEraisesectionheading{\section{Introduction}}

\IEEEPARstart{I}{n} today's interconnected world, all types of private and proprietary information---from personal health records and communications, to government records, bank information, and intellectual property---are accessible over the Internet. While the benefits of nearly ubiquitous, on-demand access to information are significant, equally significant are the risks that such access poses to individuals and organizations by exposing them to cyber attacks. The risks posed by cyber threat include financial losses and political instability, as  demonstrated by  high-profile attacks, including massive Equifax and Yahoo! data breaches and Wannacry ransomware, which paralyzed critical infrastructure worldwide to include hospitals in the US and UK. For society to continue enjoying the benefits of an open, worldwide Internet, it is critical that we tame the rapidly growing cyber threats posed by a variety of state and non-state actors. 

One approach to combating cyber threats is to develop technologies that anticipate them before an actual cyber attack occurs. 
The intuition behind this forecasting approach is the following. Cyber attacks do not occur in a vacuum. To conduct a cyber attack, hackers first have to choose a target, identify the attack surface (i.e, vulnerabilities in the target's software and hardware infrastructure), acquire the necessary exploits, malware and expertise to use them, and potentially recruit other participants. Other actors---system administrators, security analysts, and even victims---may discuss vulnerabilities or coordinate a response to exploits. These activities are often conducted online, leaving a variety of digital traces that can be mined to extract signals of  pending  attacks well before suspicious activity is noted on the target system.  

Identifying useful signals of  impending cyber attacks poses several research challenges. 
First, while some of the data relating to activities of cyber actors is openly available, malicious actors often obfuscate their actions using anonymized and encrypted Internet protocols. Second, the behavioral processes generating activities of interest are likely to be weak, sparse, and transient, posing significant challenges to picking them out from among massive quantities of entirely innocuous activity. Finally, 
translating the signals 
to generate a warning about 
a cyber attack presents yet another challenge.

Under the IARPA-funded CAUSE program, USC Information Sciences Institute has  developed  an end-to-end prototype, called EFFECT, to  forecast emerging cyber threats. 
This paper  describes two machine learning methods for time series  prediction that 
are used by EFFECT to 
forecast cyber attacks. The methods take as input historical data 
to learn a model of cyber attacks. 
These models capture  patterns present in historical data that help forecast new cyber attacks. We show that we can improve the predictions of these baseline models by leveraging signals from external Web data sources. 

To construct external signals, EFFECT harvests data from a variety of sources, including vulnerability databases, malicious email and malware trackers, but also from sources not conventionally used in security applications, such as social media, blogs and darkweb forums. From these data sources we extract a variety of time series, each representing the number of daily occurrences of  cyber security-related terms. The  time series are used as external signals in the forecasting task.

To train the forecasting models, and to evaluate their predictions, we use the ground truth data about  cyber attacks provided through the CAUSE program from two companies.  The ground truth data comprise of attacks intercepted at both organizations, which correspond to three types of events: malware installed on user's computer (endpoint-malware), malicious email (malicious-email) and malicious destination a user navigates to (malicious-destination).

Specifically, our paper makes the following contributions: 
\begin{itemize}
\item Describes and evaluates a time series forecasting method based on autoregressive models that leverage external time series in the prediction task
\item Describes and evaluates a time series forecasting method based on a neural network.
\item Identifies signals from online data sources that  consistently improve predictions of cyber attacks in the ground truth data.
\end{itemize}

The rest of the paper is organized as follows. We first describe the Web data sources used by EFFECT. Next, we describe in detail the two machine learning algorithms used in the forecasting task, as well as how they are trained. Finally, we evaluate prediction results and discuss the significance of predictive signals identified by the system.
\section{Data}
\label{sec:datasets}
In this section, we briefly describe the Web data sources used by EFFECT, how signals are extracted from these sources, and the ground truth data used to train and validate forecasting models.

\subsection{Web Data Sources}
\subsubsection{Dark/deep web}  Deep and Dark (D2Web) web are non-indexed sites on the open Internet which are accessed using anonymization protocols (most notably the TOR protocol). These web sites host discussion forums and marketplaces, which  are often used for malicious or illicit purposes, for example, to buy and sell drugs,  guns, hacked data or exploits. It has been shown that the activity on these websites can signal potential cyber attacks \cite{7745465, 7745437,tavabi2018dark, sapienza2017early}. The infrastructure used to collect the data is described in ~\cite{robertson2017darkweb,nunes2016darknet}. 
Close to 300 different D2Web sites were used for this study.

\subsubsection{Twitter} It has been shown that discussions on social media can be used as signals for detecting cyber threats \cite{sapienza2017early, Sabottke:2015:VDA:2831143.2831209}. Therefore we collected tweets which where either posted by security experts or contained cyber security related keywords by manually compiling a list of almost 250 experts and 1500 multilingual security related terms.

\subsubsection{Blogs} Security blogs are posts written by expert analysts regarding news and events in the cyber security domain at the time of writing. The EFFECT system is crawling about 70 different websites to collect blog posts. Using security blogs as a data source for predicting cyber events is relatively new and was originally proposed in \cite{tavabi2018dark,sapienza2017early}. 


\subsubsection{Vulnerability Database} Software vulnerabilities are often exploited by malicious actors in cyber attacks~\cite{martin2017effective,Tavabi2018iaai}. National Vulnerability Database (NVD) is the largest publicly available repository which contains information about reported vulnerabilities in software. We collected vulnerabilities for different software products to evaluate their role in predicting cyber attacks. 

\subsubsection{Honeypots}
A honeypot is a security resource with the goal of having a system probed and attacked. Traffic reaching honeypots may be malicious and can provide a window into hacker activities. We collected data from a network of ten honeypots deployed by the EFFECT team~\cite{GoogleDorks}: specifically, the number of queries received daily by each honeypot serves as an external signal in the cyber forecasting task. The honeypots were deployed and data was collected starting in October 2017.



\subsection{External Signals}
Data sources already providing the number of daily events of a specific type were used as is. 
To use the textual data collected by EFFECT for the prediction task, we compiled a list of 50 important keywords in the cyber security domain. 
These keywords included terms, such as \textit{0day}, \textit{exploits}, \textit{vpn}, and \textit{vulnerabilities}. The full list of terms can be found in the Supplementary Information (SI). 
We created external signals from the data sources by extracting the time series of the number of daily occurrences of each cyber term, giving us 50 external signals for each of the D2Web, social media and blogs domains.

\subsection{Ground Truth}
We use as ground truth (GT) data about three types of cyber attacks provided by  two organizations to the CAUSE program (refered to as OrgA and OrgB). The data contains occurrence times of three types of cyber attacks:
\begin{itemize}
\item{An endpoint-malware} attack is recorded when anti-virus software used by the organization finds malware installed on end-user's system.
\item{A malicious-email} attack is the receipt of an email that contains a malicious email attachment and/or a link to a known malicious destination. 
\item{A malicious-destination} attack is recorded when end-user clicks on a malicious URL.
\end{itemize}

These data cover a period from July 2017 to January 2018.

\section{Methods}\label{sec:methodology}
In this section, we describe the time series modeling approaches we use for the problem of cyber attack prediction.  We give an overview of classical approaches based on autoregressive models and more recent approaches which use neural networks for prediction. 
Next, we describe the training and predicting methods for the two machine learning approaches used in forecasting cyber attacks.

\subsection{Forecasting Task}

\begin{figure}[h!]
  \centering
  \includegraphics[width=0.5\textwidth]{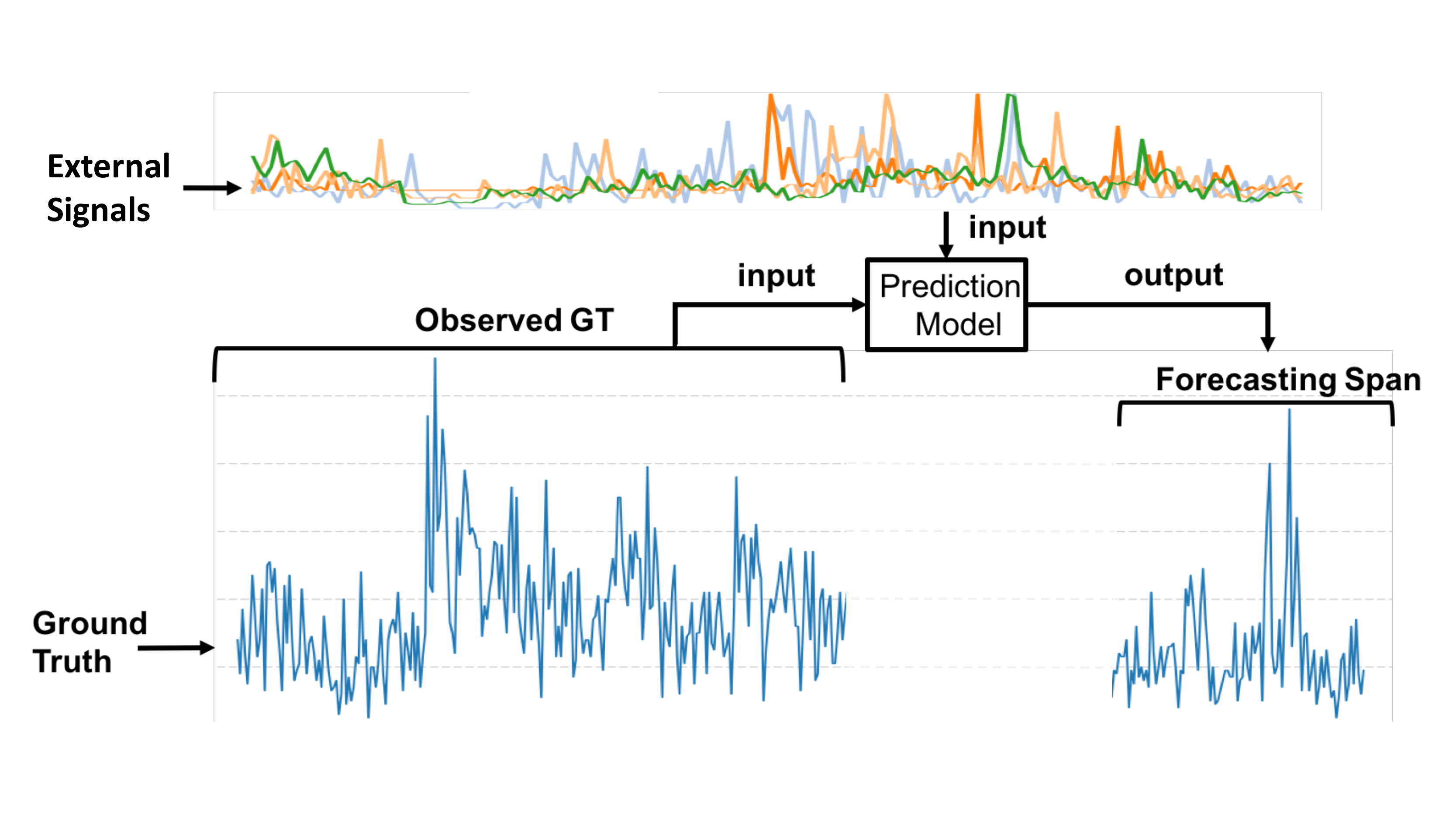}
  \caption{Illustration of cyber attack forecasting. We assume that ground truth data representing historical cyber attacks is provided for training prediction models, along with external signals. The predictions are made for events occurring during the future time period. } \label{fig:illustration}
\end{figure}

Figure~\ref{fig:illustration} illustrates the forecasting task. Given a time series describing observed events in the ground truth data, our goal is to use this information, plus information from external signals, to predict new events occurring during some future forecasting time span. The prediction model is trained on the historical GT data and external signals. The illustration highlights the common case where up-to-date historical GT data may not be available, but we assume that most recent external signals are always available.

\subsection{Forecasting Models}
\subsubsection{Autoregressive Models}
\label{sec:ar-model}

We apply the popular ARIMA and ARIMAX models to the forecasting task.
ARIMA stands for autoregressive integrated moving average. The key idea behind ARIMA is that the number of current events ($y_t$) depends on the past counts and forecast errors. 
Formally, ARIMA($p$,$d$,$q$) defines an autoregressive model with $p$
autoregressive lags, $d$ difference operations, and $q$ moving average lags
(see~\cite{shumway2010time}). Given the observed series of events
$\mathcal{Y}=(y_1,y_2,\ldots,y_T)$, ARIMA($p$,$d$,$q$) applies $d$ ($\ge 0$)
difference operations to transform $\mathcal{Y}$ to a stationary series
$\mathcal{Y}^{\prime}$. Then the predicted value $y^\prime_t$ at time point
$t$ can be expressed in terms of past observed values and forecasting errors
which is as follows: 
\begin{align}
    y^\prime_t = c + \sum_{i=1}^p \alpha_i y^\prime_{t-i} + \sum_{j=1}^q \beta_j e_{t-j} + e_t\label{eq:arima}
\end{align}
Here $c$ is a constant, $\alpha_i$ is the autoregressive (AR) coefficient at lag $i$, $\beta_j$ is the moving average (MA)
coefficient at lag $j$, $e_{t-j} = y^\prime_{t-j} - \hat{y}^\prime_{t-j}$ is the forecast error at lag $j$, and $e_t$ is assumed to be the white noise ($e_t\sim \mathcal{N}(0,\sigma^2)$).  The
AR model is essentially an ARIMA model without moving average terms.

ARIMAX (Autoregressive Integrated Moving Average with Exogenous variables) is an autoregressive model that leverages (optional) external signals. In this model, the observation at a particular time point depends on immediate past observations, past forecast errors, and external variables. Like ARIMA, ARIMAX($p$,$d$,$q$) is defined with the same three autoregressive order terms as ARIMA.
Given the observed series of events $\mathcal{Y}=(y_1,y_2,\ldots,y_T)$ and optional $K$ external features\\
$\mathcal{X} = (\langle x_{11},x_{21},\ldots,x_{K1}\rangle,\allowbreak \langle x_{12},x_{22},\ldots,x_{K2}\rangle,\ldots, \langle x_{1T},x_{2T},\ldots,x_{KT}\rangle)$,\allowbreak
the model is defined as follows:
\begin{align}
    y^\prime_t = c + \sum_{i=1}^p \alpha_i y^\prime_{t-i} + \sum_{j=1}^q \beta_j e_{t-j} + \sum_{k=1}^K \gamma_k x_{kt} + e_t\label{eq:arimax}
\end{align}
Here $\mathcal{Y}^{\prime}$ is the stationary series after $d$ difference operations, $c$ is a constant, $\alpha_i$ is the autoregressive (AR) coefficient at lag $i$, $\beta_j$ is the moving average (MA)
coefficient at lag $j$, $e_{t-j} = y^\prime_{t-j} - \hat{y}^\prime_{t-j}$ is the forecast error at lag $j$, $\gamma_k$ is the coefficient for feature $x_k$ and $e_t$ is assumed to be the white noise. 

\paragraph{Training}

We use maximum likelihood estimation for learning the parameters; more
specifically, parameters are optimized with LBFGS
method~\cite{seabold2010statsmodels}. These models assume that $(p, d, q)$ are
known and the series is weakly stationary. To select the values for $(p, d, q)$
we employ grid search over the values of $(p, d, q)$ and select the one with
minimum AIC score.

\paragraph{Prediction}
For ARIMA, we use the learned parameters to estimate the event counts for the period with missing GT and then use these counts with learned parameters to predict next month's/week's cyber attack counts. For ARIMAX, similar to ARIMA, we first estimate the event counts for the period with missing GT using learned parameters, past GT, and external sources. The model then predicts next month's/week's cyber attack count with the estimated count for missing GT period and the external sources.

\subsubsection{Neural Network Models}
Neural network based models have been widely used for time series analysis as far back as 2003 with ~\cite{zhang2003time}. The autoregressive models express the predicted event count as a linear function of external signals and historical counts. However, neural network based models can capture the non-linearity by using multiple layers of non-linear activation functions. The caveat is that such models typically require a large amount of training data to accurately estimate the model parameters.

Recently, many variants of recurrent neural network units have been proposed including Long Short Term Memory (LSTM)~\cite{hochreiter1997long}, Phased LSTM~\cite{neil2016phased} and Gated Recurrent Unit (GRU)~\cite{cho2014learning}. The units are composed of various gates such as input, forget and output. The variants differ in the number, type and connections between gates.

\begin{figure}[htbp!]
  \centering
  \includegraphics[width=\columnwidth]{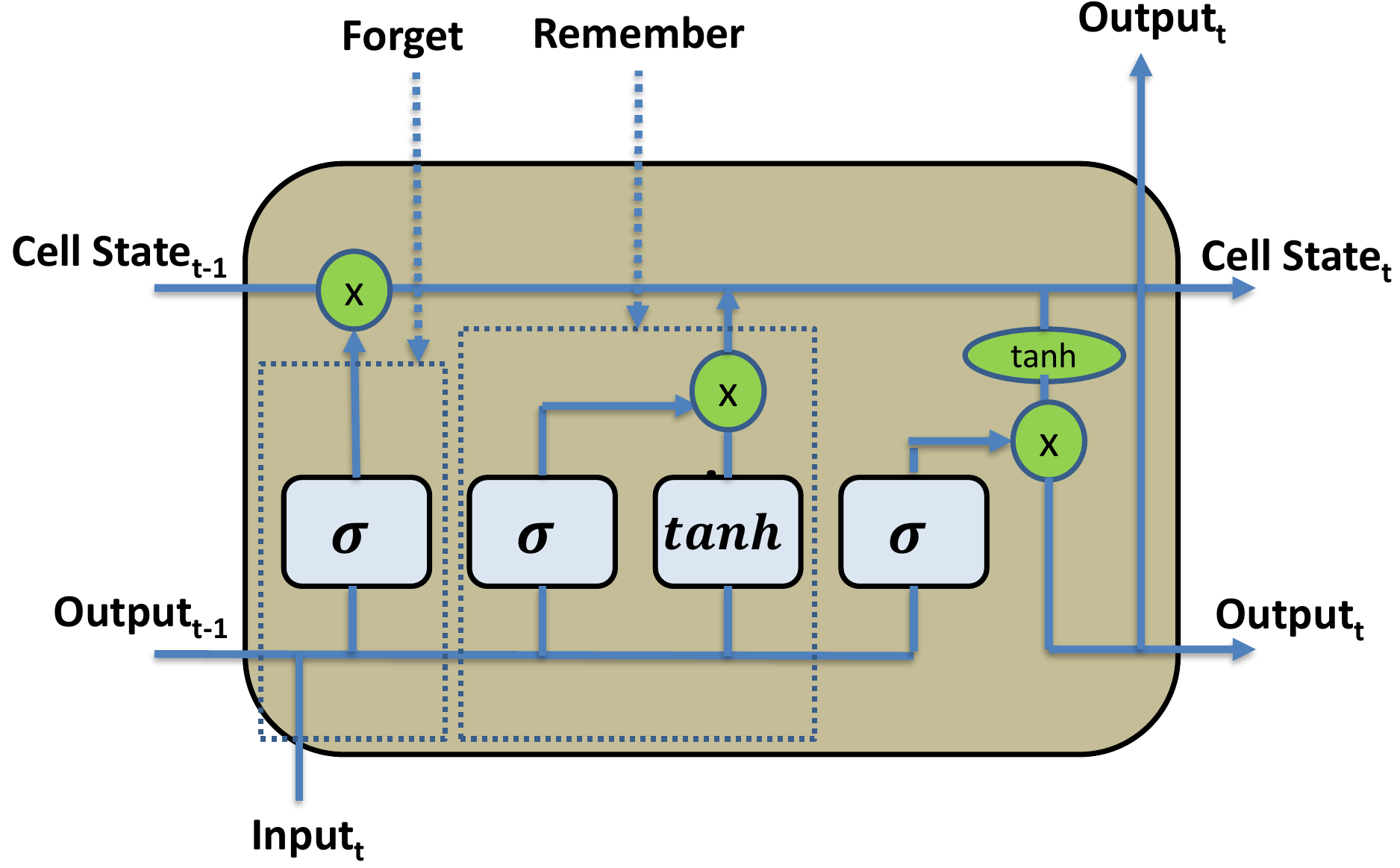}
  \caption{Long Short-Term Memory architecture. }
  \label{fig:lstm}
\end{figure}

\textit{LSTM} is composed of the above gates and a cell state, a unit is displayed in Figure \ref{fig:lstm}. The outputs of these gates are calculated as follows:
\begin{align}
f_t = \sigma(W_f [h_{t-1}, x_t] + b_f) \\
i_t = \sigma(W_i [h_{t-1}, x_t] + b_i) \\
o_t = \sigma(W_o [h_{t-1}, x_t] + b_o)\\
h_t = o_t * \tanh(C_t),
\end{align}
where $W_*$ and $b_*$ are weight matrices and bias vectors respectively. The cell state is updated using:
\begin{align}
\hat{C_t} = tanh(W_c [h_{t-1}, x_t] + b_c) \\
C_t = f_t * C_{t-1} + i_t * \hat{C_t},
\end{align}
where $C_t$ is the cell state. The parameters are learned using backpropagation and gradient descent.

\textit{PhasedLSTM} was recently developed to process irregular and sparse time series. The model extends LSTM and adds a new time gate which controls when the cell state is updated. The gate oscillates between open and closed. In open state, the LSTM cell state is updated, whereas, while in the closed state the cell state is propagated from previous time step. This gate introduces two parameters: $r_{on}$ and $\tau$ which control the ratio of duration of the open phase to full period and time period of oscillation respectively.

\textit{GRU} unit uses gates similar to LSTM unit. However, it merges the forget and input gate into an update gate which is calculated as follows:
\begin{align}
z_t = \sigma(W_z x_t + U_z h_{t-1}).
\end{align}
It also merges hidden state and output which is computed as follows:
\begin{align}
h_t = (1 - z_t) h_{t-1} + z_t \hat{h_t}\\
\hat{h_t} = tanh(W [r_t * h_{t-1}, x_t]),
\end{align}
where $r_t$ is the reset gate and is computed similar to update gate as $r_t = \sigma(W_r x_t + U_r h_{t-1})$.

\paragraph{Training} We use Adaptive Moment Estimation (Adam)~\cite{kingma2014adam} to learn the neural network parameters. We select the hyperparameters including network architecture and learning rate using cross-validation on held out data set. For monthly and weekly analyses, we use our historical data until the start of the previous month and previous week respectively.

\paragraph{Prediction} We use the learned parameters on training data to predict next month's/week's cyber attack counts. All three methods were used; however, the results for GRUs are presented as this method is computationally more efficient with similar or better performance.

\subsubsection{Baseline Models}
We compare our proposed methods against a baseline ARIMA model, which predicts the number of future events from a Poisson distribution with rate $\lambda$ as the average number of past events over a time window $W$. Formally,
\begin{align}
    \lambda_t = \frac{1}{W}\sum_{i=1}^W y_{t-i}\nonumber\\
    y_t \sim \text{Poisson}(\lambda_t),
\end{align}

where $W$ is selected as the number of time units in a training period.  


\subsection{Evaluation Metrics}\label{sec:ts-metrics}
We use two different measures for quantitative evaluation of predictions made by the models.

First, we use program-wide metrics to evaluate the accuracy of predictions. The metrics work by matching predictions against the ground truth data. To that end, we convert predicted event counts for each day to warnings of attacks predicted for that day.  
We then match the warnings against ground truth data 
using the Hungarian matching algorithm \cite{kuhn1955hungarian}.
The algorithm compares predicted warnings $w$ to ground truth events $g$. The algorithm identifies the mutually exclusive pairs $M=\{(w,g)\}$, such that the sum of similarities $\sum_{(w,g)\in M}sim(w,g)$ is maximized. 
If $w$  occurs within some time window of event $g$, then $sim(w,g)$  equals the quality score. Otherwise $sim(w,g) = 0$.  
The window around the actual events varies based on the event type: for endpoint-malware, it is 0.875 days, malicious-destination within 1.625 days and malicious-email within 1.375 days.
Using the matching algorithm, we can consistently quantify the precision and recall of the predictions, and calculate the resulting F1-score, which gives the geometric mean of precision and recall.

\begin{itemize}

\item \textit{Precision}
\begin{align*}
	\text{P} = \frac{TP}{TP+FP} 
\end{align*}

\item \textit{Recall}
\begin{align*}
	\text{R} = \frac{TP}{TP+FN} 
\end{align*}

\item \textit{F1}
\begin{align*}
	\text{F1} = \frac{2*P*R}{P+R}
\end{align*}
\end{itemize}
Here, $TP$, $TN$ and $FN$ denote the true positives, true negatives and false negatives respectively.

To measure the forecasting error of the model, we use three measures: (a) mean absolute error (MAE), b) root mean squared error (RMSE), and c)
mean absolute scaled error (MASE)~\cite{hyndman2006another}. These measures are
defined in terms of forecasting error, $e_t = y_t - y\prime_t$, at daily time steps $t$, where $y_t$ and $y\prime_t$ are the true and predicted values,
respectively.

\begin{itemize}
\item \textit{Mean Absolute Error}
\begin{align*}
	\text{MAE} = \frac{1}{T} \sum_{t=1}^T|e_t|
\end{align*}

\item \textit{Root Mean Squared Error}
\begin{align*}
	\text{RMSE} = \sqrt{\frac{1}{T} \sum_{t=1}^T|e_t|^2}
\end{align*}

\item \textit{Mean Absolute Scaled Error}
\begin{align*}
	\text{MASE} = \frac{\frac{1}{T} \sum_{t=1}^T|e_t|}{
		\frac{1}{T-1} \sum_{t=2}^T|y_t - y_{t-1}|
		}
\end{align*}
\end{itemize}

\section{Results}

\subsection{Baseline Monthly vs Weekly Analysis}
\begin{figure}[h!]

    \includegraphics[width=\columnwidth]{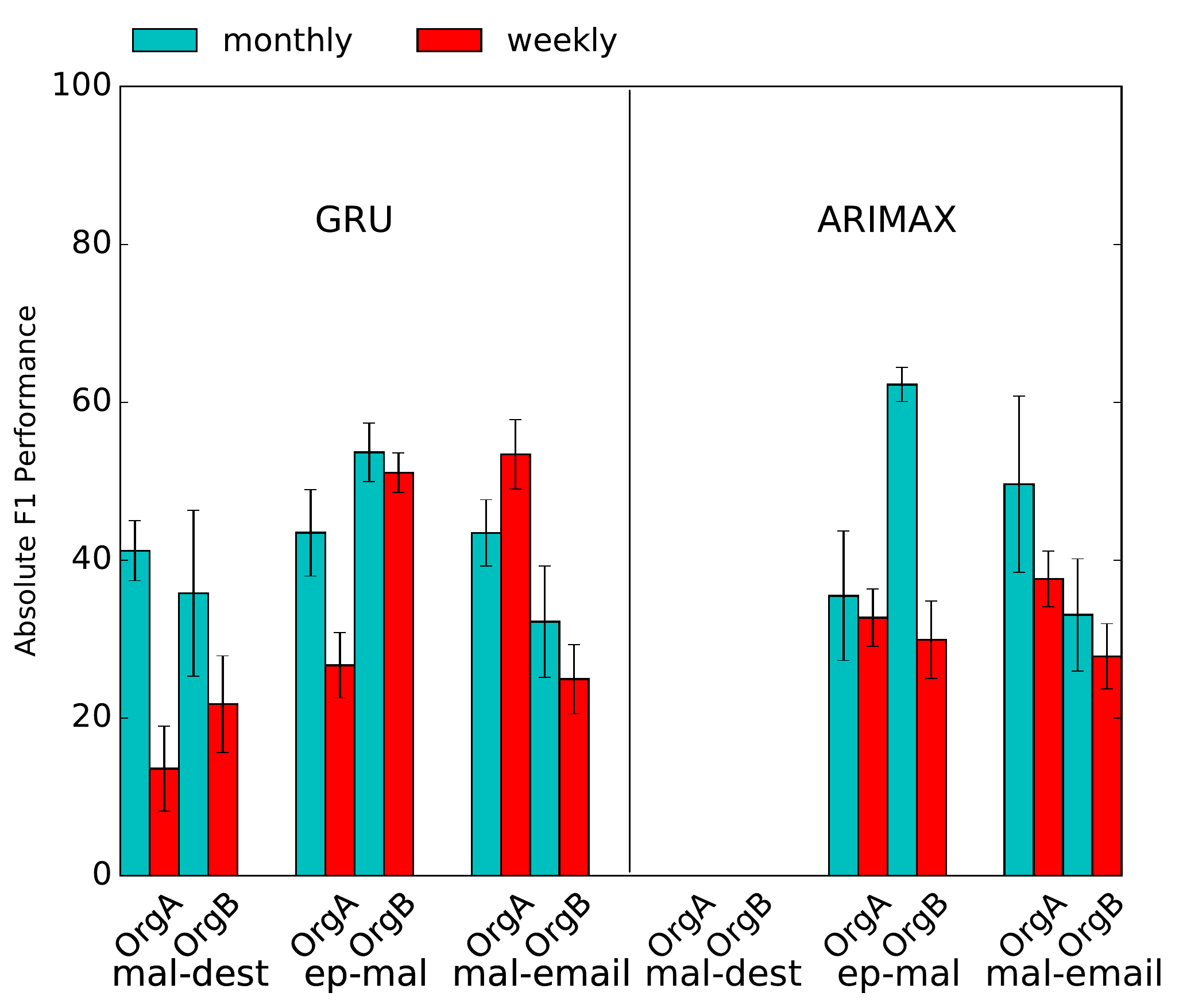}
    \caption{Baseline Monthly vs Weekly Predictions}

  \label{fig:monthly-weekly}
\end{figure}

In order to understand how the granularity of the prediction window affects performance we test two prediction windows: (i) month, and (ii) week. For monthly prediction, all historical GT data and external signals through the previous month are used to make predictions for the next month. For weekly prediction, GT data up to the start of the previous month and external signals through the previous week are used to make next week's predictions. We use this framework, because GT data is released on a monthly basis within the CAUSE program, while external signals are available on a continual basis. 

\begin{figure}[h!]
	\includegraphics[width=\columnwidth]{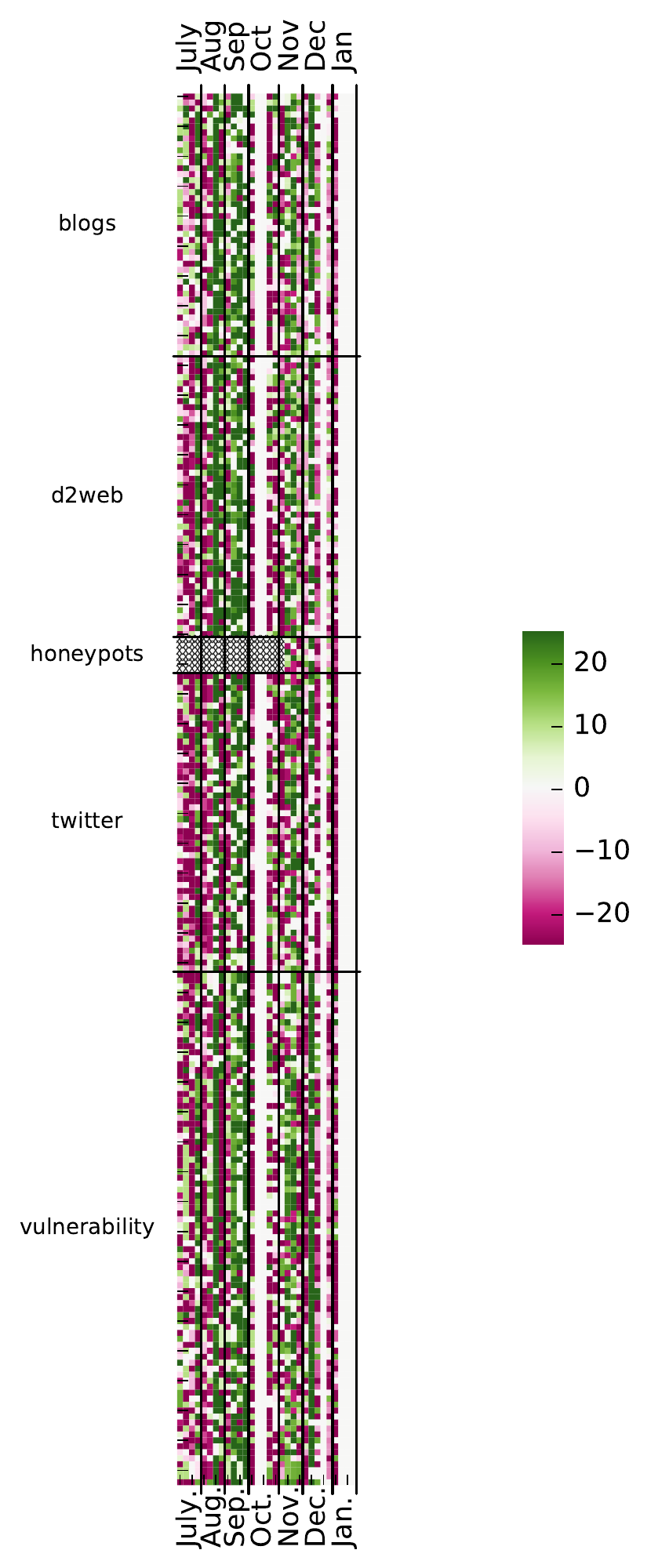}
	\caption{Difference in Weekly GRU F1 Performance between signal and baseline - OrgA Endpoint Malware}
    \label{fig:gru-a-malware-temporal}
\end{figure}

Figure \ref{fig:monthly-weekly} shows prediction performance of the baseline models, which use historical GT data through the end of the previous month only, as a function of time for the two target organizations and three event types. Information contained in the historical data allows models to achieve decent prediction performance, except for the malicious-destination event type, which contains too few events for ARIMAX to learn from. In contrast, the GRU model is able to learn from even sparse data. With respect to the granularity of prediction, making predictions for a week should yield higher performance than monthly predictions. However, here we observe that there is either comparable performance, with no statistical difference between monthly and weekly, or decreased performance from monthly to weekly. This is in part due to the sparsity of the data, as removing any weeks with zero events (and thus zero F1) raises the weekly score to be comparable to the monthly in most cases. This is also in part due to the fact that when we train on weeks, we may experience something similar to Fig.~\ref{fig:gru-a-malware-temporal}, where we see strong performance in one or two weeks of a month get washed out by weak performance during the rest of the month. Considering these aspects, monthly performance is used an the evaluation time frame in our analyses. We only consider monthly predictions for the rest of the paper and results for weekly performance can be found at the end of the document. 


\subsection{Finding Correlated Signals}
\begin{figure}[h!]
  \includegraphics[width=\columnwidth]{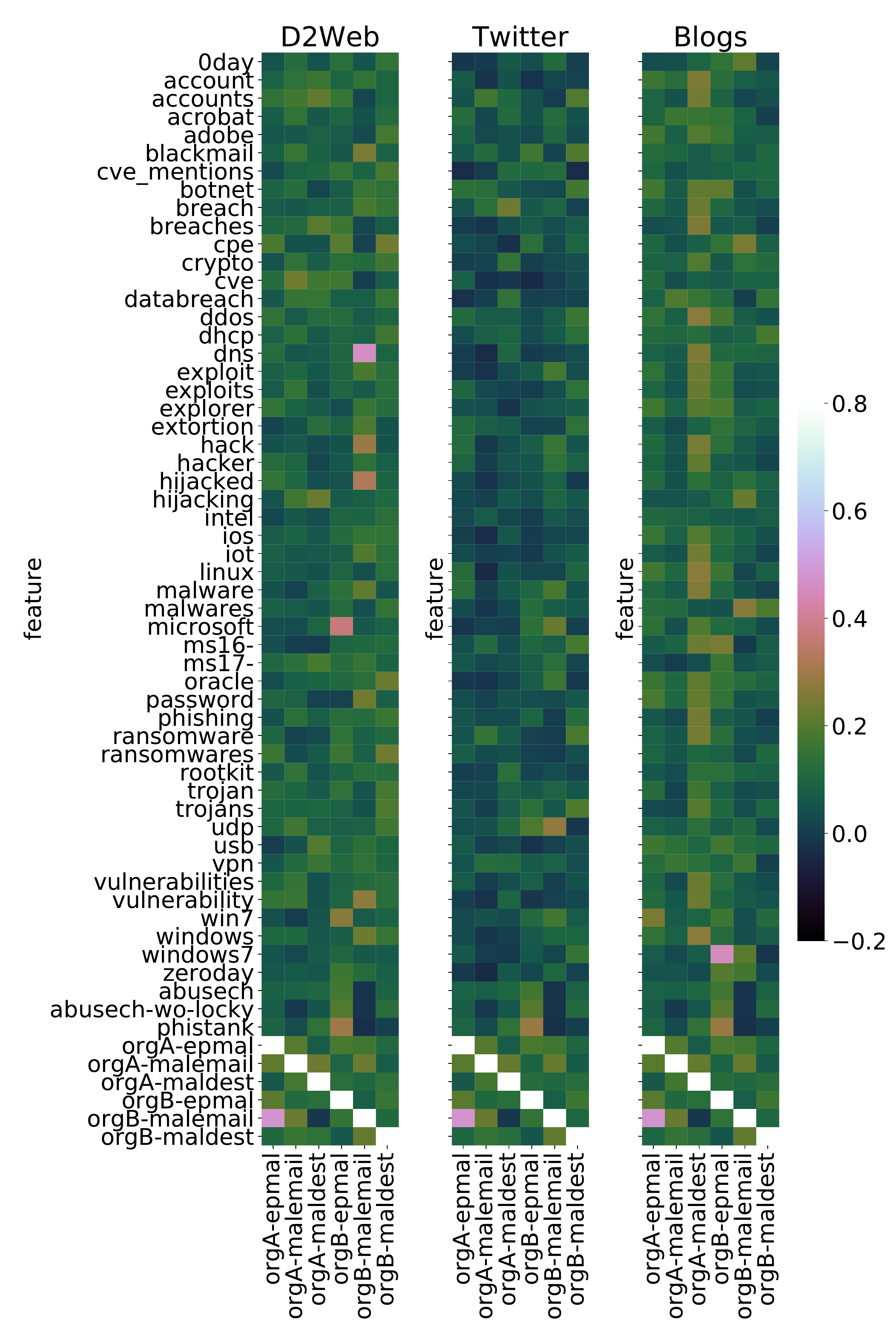}
  \caption{Correlation of word count signals with ground truth}
  \label{fig:wordcountCorrelations}
\end{figure}

Correlation analysis is done as a pre-processing step to pick out signals that may have predictive value. 
For each target time series, we compute the lagged cross correlation with all other signals. The lagged signal is created by shifting the time series by a certain amount. We used lags from $-30$ to $0$ days and chose the lag with highest correlation. Figure ~\ref{fig:wordcountCorrelations} shows the correlation of three data sources with ground truth data. For OrgA, Blogs overall have the highest correlations followed by D2Web signals and then Twitter. For OrgB, D2Web has the highest correlations, with Blogs following and then Twitter. In both cases, Twitter is the lowest. OrgB endpoint malware has the highest correlation with any of the external signals and \textit{oracle}, \textit{accounts}, \textit{blackmail} and \textit{malwares} are the keywords with highest average correlations. The same analysis was done for the other two data sources, vulnerabilities and honeypots. Refer to the Appendix section for their corresponding plots. Out of these, vulnerabilities related to f5 big-ip and Oracle have high correlation with malicious email, especially orgB. Oracle vulnerabilities also have high correlations with malware endpoints as do Mozilla and Novell. 

\subsection{Identifying Predictive Signals}
From the correlation analysis we have 285 signals we can use in our models. As these 285 signals span several data sources and are individually evaluated on several different prediction tasks, it is difficult to reason about any performance measure without summarizing it visually. We analyze the performance of each signal within a data source and then provide comparison across various data sources.

Before using external signals in prediction, we first align them using correlation analysis with GT: we determine the lag where
maximum correlation occurred between a temporal feature and GT, and use the
lag for alignment. For ARIMAX we set a maximum auto-regressive lag of 7 days. In each section, we report the five signals with the highest average lift (defined as the ratio of the model trained with signal X to the baseline model) in F1 score for each event-type-target combination. Summary results using forecasting error measures, specifically, RMSE are in the supplemental material in Table \ref{tab:perf-monthly-orga-gru} to Table \ref{tab:perf-weekly-orga-gru}.

\subsubsection{Blog Signals}
\begin{figure*}[h!]
\centering
  \includegraphics[width=1.5\columnwidth]{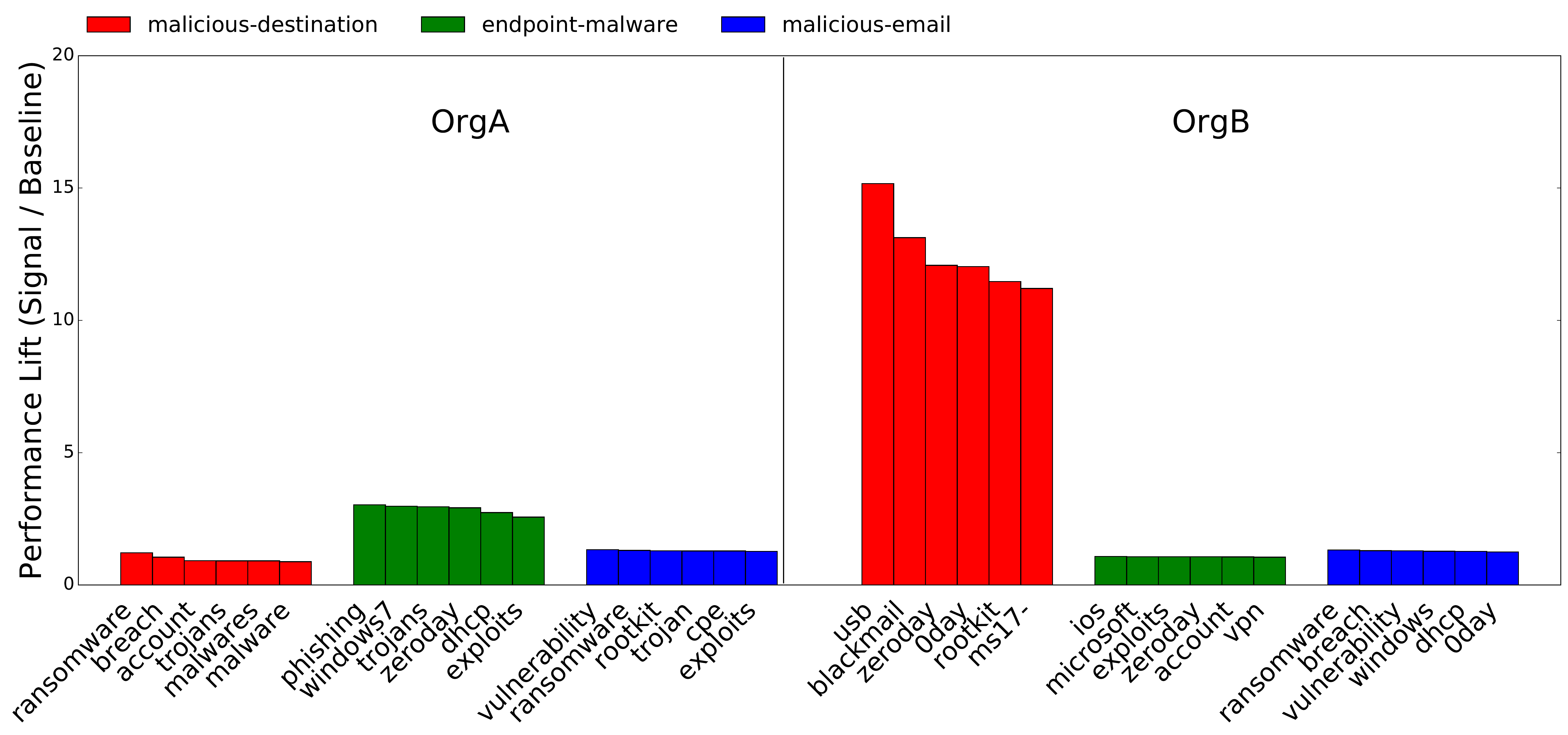}
  \caption{Monthly GRU F1 performance of blog signals. }
  \label{fig:gru-blogs}
\end{figure*}
Figure \ref{fig:gru-blogs} shows improvements in prediction performance due to signals from blogs, evaluated using GRUs. The results using ARIMAX can be found in the supplemental material. For OrgA the top signals work best for endpoint-malware and for OrgB, the top signals work best with the malicious-destination events. Additionally, OrgB's malicious-destination events can be better predicted than other event from either organization, with \textit{usb}, \textit{blackmail} and \textit{zeroday} being the best keywords. For OrgA, endpoint-malware was the most predictive with keyword is \textit{phishing}. For malicious-email, it is very difficult for blogs to predict this event type well with neither organization above a lift of 2. However, \textit{vulnerability} and \textit{ransomware} were both in the top 3 signals for both organizations.  Interestingly, the majority of the keywords that were the best for OrgA were not best for OrgB. The only terms that carried over were \textit{ransomware}, \textit{vulnerability} and \textit{zeroday}. The ground truth events and associated best signals are different for the two companies analyzed. This is indicated in the correlation analysis above and Figure \ref{fig:gru-blogs}. 

\subsubsection{D2Web Signals}
\begin{figure*}[h!]
\centering
  \includegraphics[width=1.5\columnwidth]{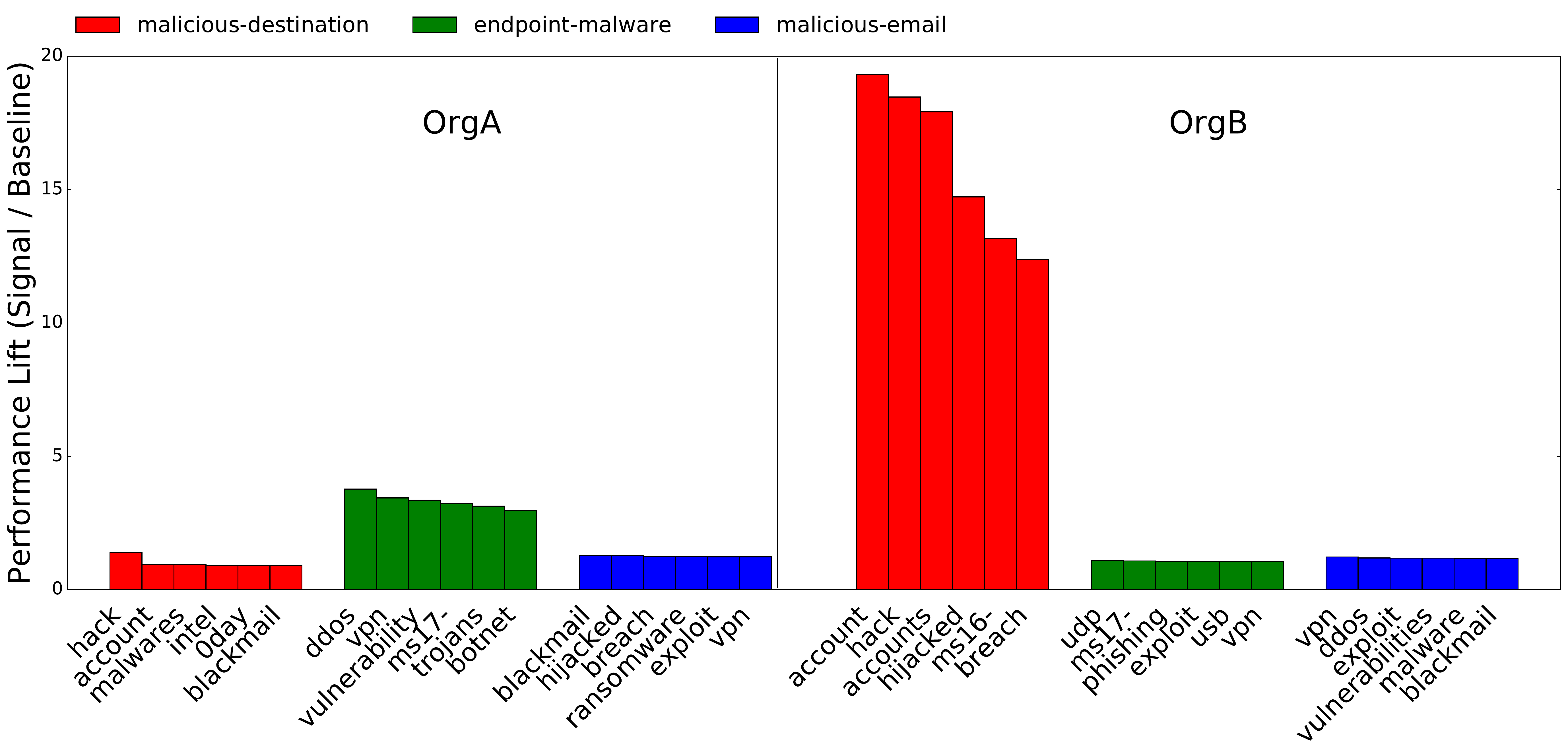}
  \caption{Monthly GRU F1 performance of D2web signals.}
  \label{fig:gru-d2web}
\end{figure*}
Figure \ref{fig:gru-d2web} highlights the results of key terms in D2Web posts for GRU and this performance is better than ARIMAX predictions. Similar to GRU blogs, predictive signals work best with malicious-destination, especially for OrgB. This may indicate that blogs and d2web as a source are very similar in their underlying signals. Other similarities to blogs are how key term signals perform better than the baseline for endpoint-malware works better for OrgA than for OrgB. The keywords overlap more than for blogs, but still not a significant amount. However, the top two keywords for malicious-destination for both organizations were \textit{account} and \textit{hack}.

\subsubsection{Twitter Signals}
\begin{figure*}[h!]
\centering
  \includegraphics[width=1.5\columnwidth]{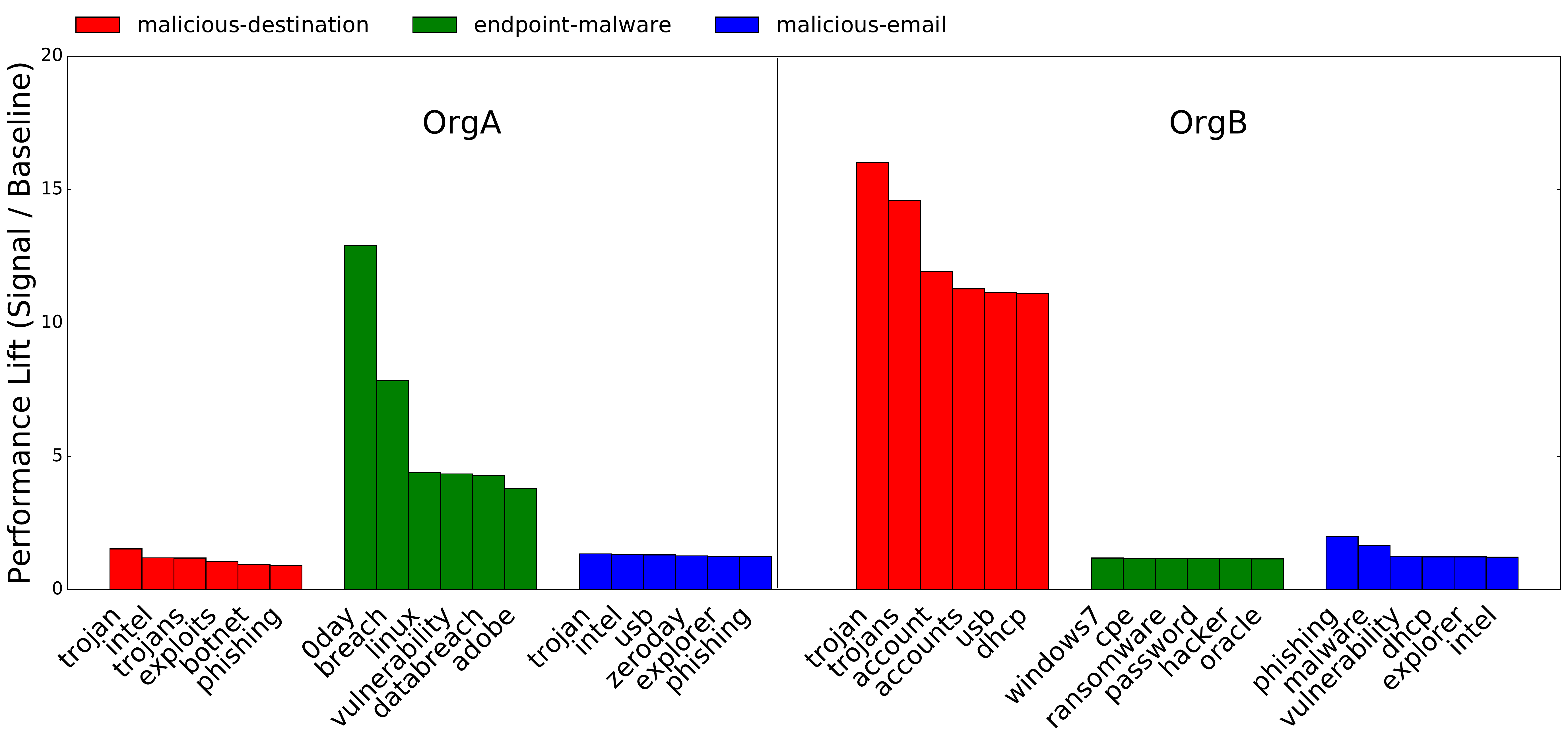}
  \caption{Monthly GRU F1 performance of Twitter signals.}
  \label{fig:gru-twitter}
\end{figure*}

Figure \ref{fig:gru-twitter} illustrates the performance of Twitter keywords as external signals for cyber attack prediction using GRU. Similar to blogs and D2Web, malicious-destination events for OrgB achieve higher lift using the external signals than other event types with endpoint-malware being a close second for OrgA. Extremely low improvement over baseline for malicious email suggests that keyword counts on Twitter do not provide information predictive of malicious email counts which is the same for blogs and D2web. For malicious destination, the terms \textit{trojan}, \textit{trojans}, \textit{account} and \textit{accounts} are in the top 5 predictive signals for OrgB.  The unique aspect of the Twitter signal is how the endpoint-malware for OrgA and the malicious-destination for OrgB using GRUs has more of a gradient than blogs or D2Web. As such, \textit{0day} has a  significantly more comparative advantage over the second best keyword of \textit{breach} for OrgA endpoint-malware.


\subsubsection{Vulnerability Signals}
\begin{figure*}[h!]
\centering
  \includegraphics[width=1.5\columnwidth]{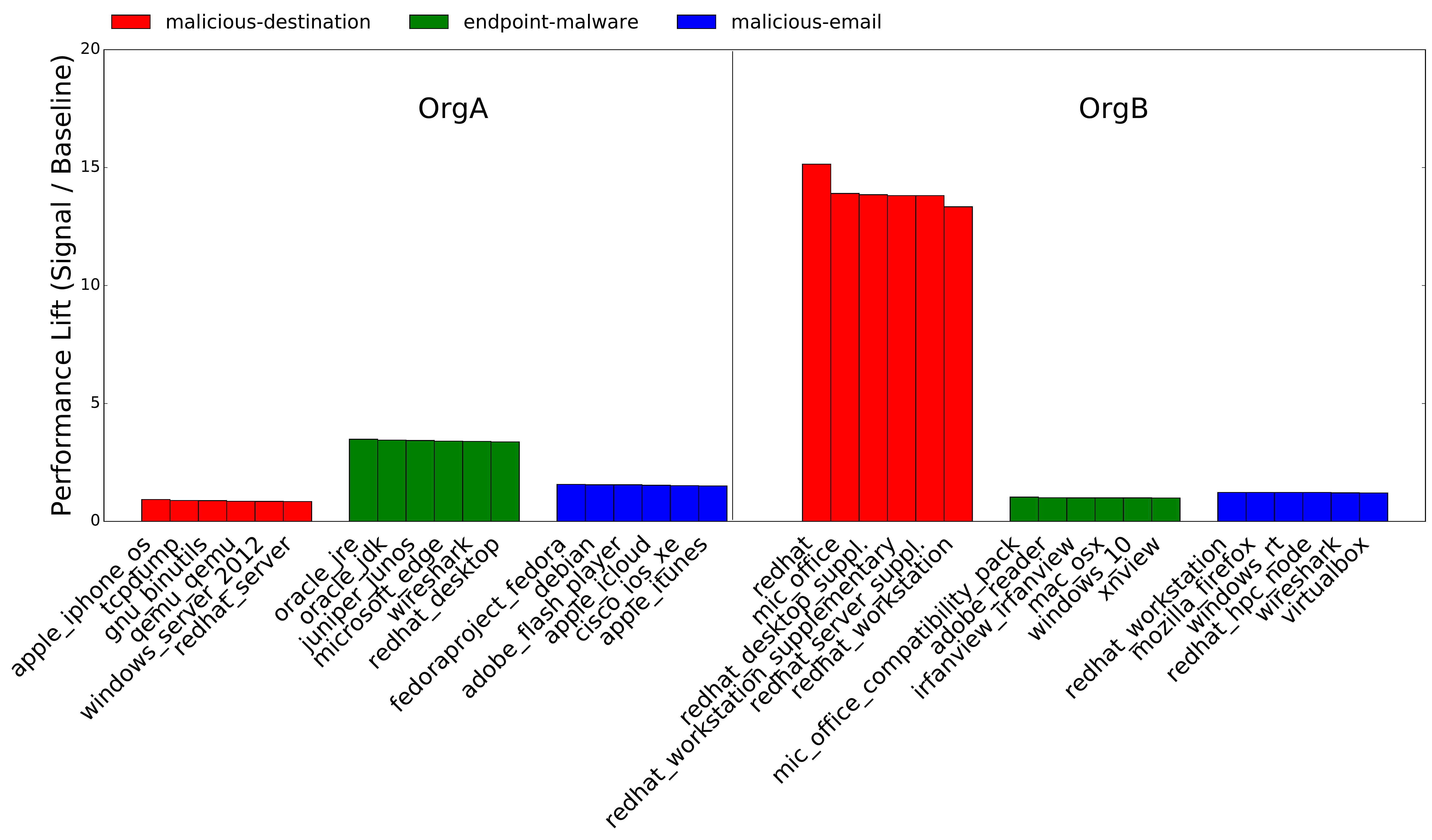}
  \caption{Monthly GRU F1 performance of vulnerability signals.}
  \label{fig:gru-vulnerability}
\end{figure*}

The predictive capacity of published vulnerabilities in software is illustrated in Figure \ref{fig:gru-vulnerability}. The overall performance gain is similar to other sources for the GRU models. Here we observe that Redhat vulnerabilities are of predictive improvement on malicious-destination attacks for OrgB which is counter to what we would expect since both the companies use Microsoft products. In general, tracking vulnerabilities can help identify susceptibility of the organizations to attacks but only for one event type per organization.  We observe that malicious emails continue to prove challenging to predict using such signals. In this case, it is logical as malicious emails do not require exploiting software vulnerabilities. Similar to other signals, ARIMAX does not leverage external signals well (see Figure \ref{fig:arimax-vulnerability} in Appendix) with maximum improvement about 200\% of baseline. This is due to how GRU handles sparsity in a signal far better than ARIMAX and vulnerabilities are the most sparse source. 

\subsubsection{Honeypot Signals}
\begin{figure*}[h!]
\centering
  \includegraphics[width=1.5\columnwidth]{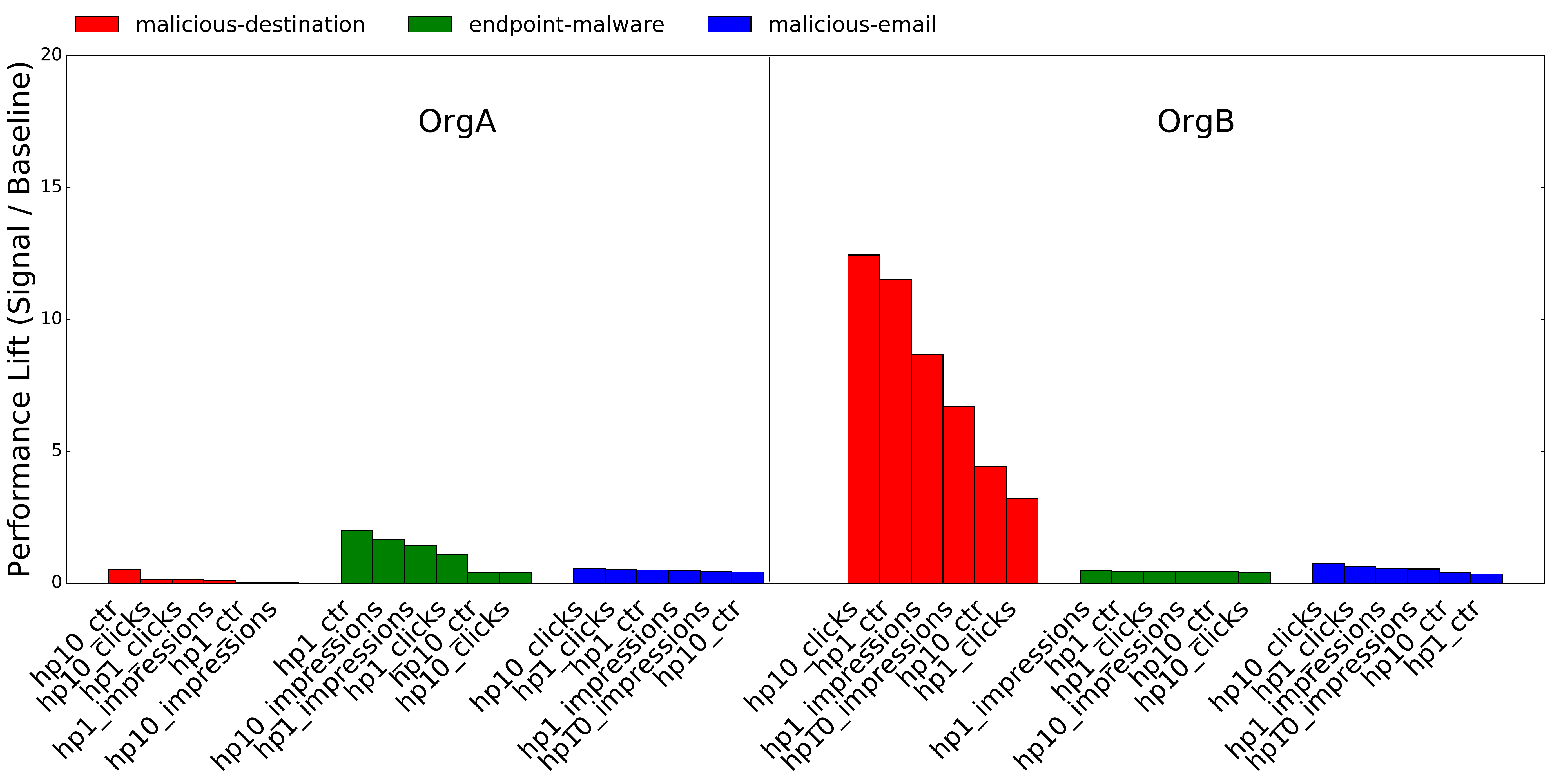}
  \caption{Monthly GRU F1 performance of honeypots signals.}
  \label{fig:gru-honeypots}
\end{figure*}
Fig.~\ref{fig:gru-honeypots} shows the common theme that malicious-destination, particularly for OrgB is the best to predict using GRUs. It is clear that this may be the case due to poor baseline performance on malicious-destination; any performance gain over baseline does not have to be substantial to achieve substantial lift. Additionally, the GRU model works better than ARIMAX, again due to the sparsity of the data. Furthermore, the honeypot signal has the least predictive power for all the sources considered. This is evident that the best honeypot signal had a lift around 11 whereas  all  other sources had best signals well above a lift of 15. This can be explained however by the fact that we did not start collecting honeypot data until the end of October 2017, missing key activity between July and October.

\subsection{Selecting the Best Signals Overall}
\begin{figure}[h!]
	\includegraphics[width=\columnwidth]{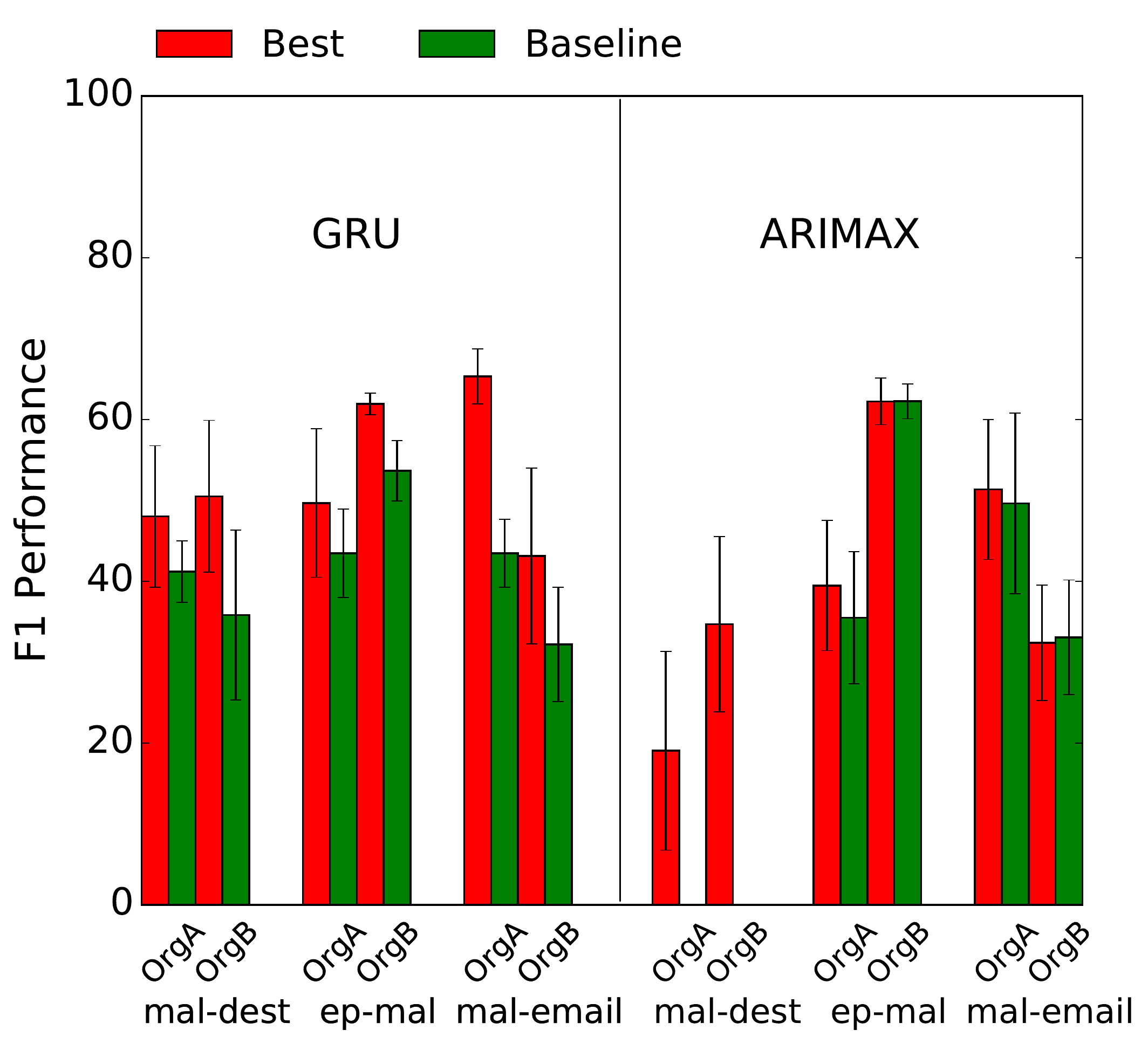}
	\caption{Comparing monthly performance models trained on the best signal for each configuration against baseline.} 
    \label{fig:best-sig-vs-baseline}
\end{figure}

In order to assess how well the relative lift identifies predictive signals, we consider the signals with the highest relative lift for any particular event-type-target configuration and compare its absolute F1 performance against baseline. Fig.~\ref{fig:best-sig-vs-baseline} illustrates that when we choose the best signal accordingly, we see that the model on average outperforms the baseline every time. In Table ~\ref{tab:best-sigs} we see that these signals with the highest absolute F1 score come predominantly from vulnerabilities, with blogs and D2Web following in contributions. Additional plots of best performing monthly and weekly signals identified by both models can be found in Supplementary Information (SI) file.  


In Fig.~\ref{fig:top-signals-temporal} we observe substantial variance in the density of predictions made across the evaluation period, from GRU trained on \textit{d2web\_zeroday} having made predictions each week to ARIMAX trained on \textit{d2web\_zeroday} not having substantial information to make predictions past the first week. Interestingly, for ARIMAX malicious-email trained on OrgB, the \textit{baseline\_arima} model seems to perform the best. 

The consistency among best performing signals in monthly forecasts suggests the methods work well to identify useful signals (see SI for details and comparison to weekly forecasts). For malicious destination-type event,   \textit{twitter\_trojan}, \textit{d2web\_hack} and \textit{d2web\_account} work well across both targets with both models. To forecast malware-type events, the best signals are \textit{twitter\_0day}, \textit{twitter\_breach}, \textit{twitter\_vulnerability} for OrgA and \textit{twitter\_windows7}, \textit{twitter\_cpe} and \textit{twitter\_ransomware} for OrgB. These choices make sense, as CPE (Common Product Enumeration) numbers identify vulnerable software. Interestingly, for malicious email events, the vulnerabilities data source provides best signals for OrgA, while for OrgB, the best performing signals are \textit{twitter\_phishing}, \textit{twitter\_malware}, and \textit{blogs\_ransomware}.   Ransomware is a type of malware that is usually spread through email.

\begin{figure}[h!]
\centering
\includegraphics[width=\columnwidth]{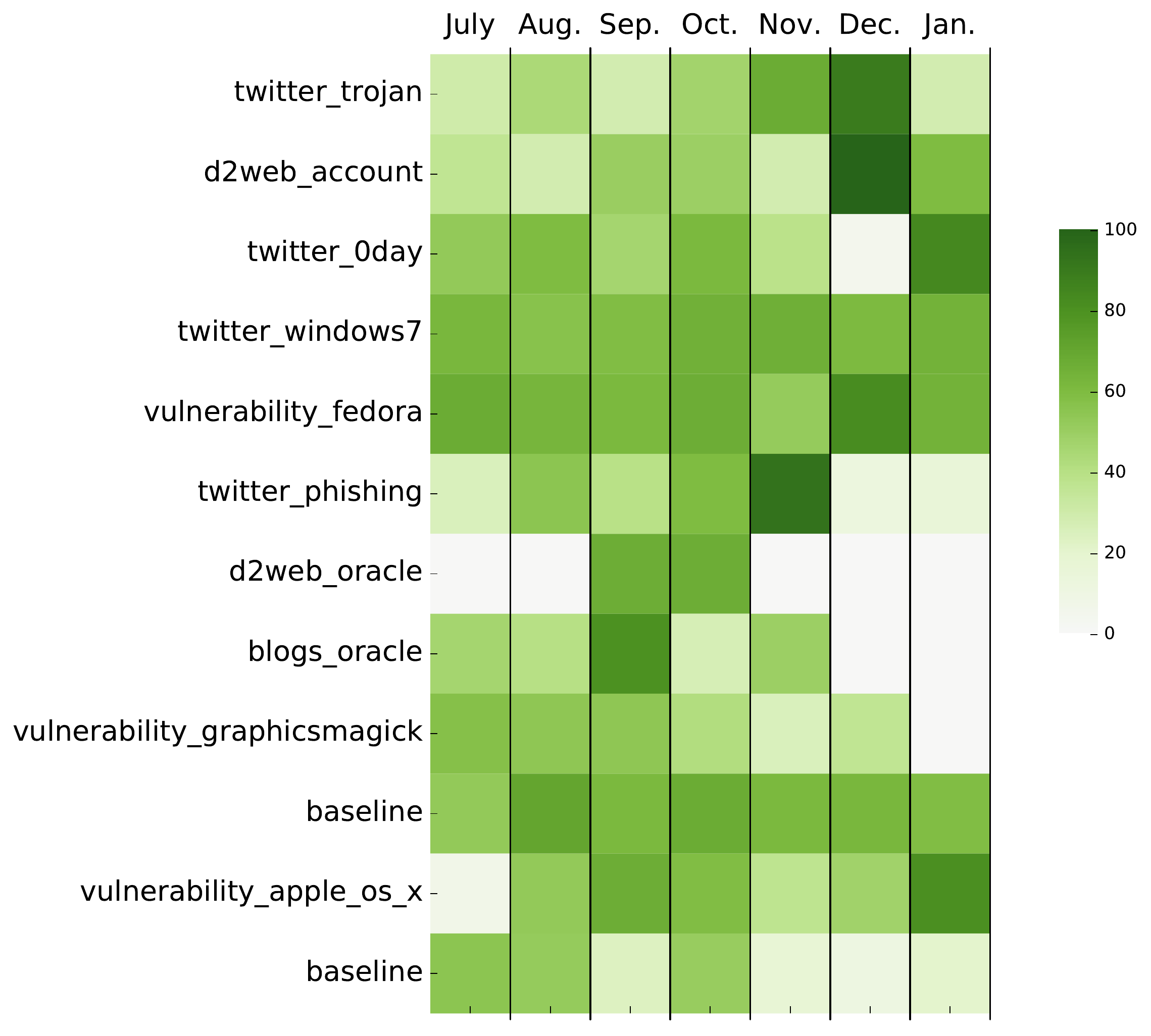}
\caption{Temporal performance of the best  signals.}
\label{fig:top-signals-temporal}
\end{figure}

\begin{table}[!t]
\centering
\caption{Best performing signals for each configuration.}
\label{tab:best-sigs} 
\begin{tabular}{llccc}%
\hline%
\textbf{Model}&\textbf{Event\_Type}&\textbf{Org}&\textbf{Signal}&\textbf{F1}\\%
\hline%
GRU	&	mal-dest.	&	A	&	twitter\_trojan	&	48.00\\%
GRU	&	mal-dest.	&	B	&	d2web\_account	&	50.50\\%
GRU	&	ep-malware	&	A	&	twitter\_0day	&	49.66\\%
GRU	&	ep-malware	&	B	&	twitter\_windows7	&	61.91\\%
GRU	&	mal-email	&	A	&	vuln.\_fedoraproj\_fedora	&	65.30\\%
GRU	&	mal-email	&	B	&	twitter\_phishing	&	43.12\\%
\hline%
ARIMAX	&	mal-dest.	&	A	&	d2web\_oracle	&	19.05\\%
ARIMAX	&	mal-dest.	&	B	&	blogs\_oracle	&	34.69\\%
ARIMAX	&	ep-malware	&	A	&	vuln.\_graphicsmagick	& 38.46\\%
ARIMAX	&	ep-malware	&	B	&	baseline    &	62.22\\%
ARIMAX	&	mal-email	&	A	&	vuln.\_apple\_os\_x	&	51.34\\%
ARIMAX	&	mal-email	&	B	&	baseline	&	32.40\\%
\hline%
\end{tabular}%
\end{table}

\section{Related Work}
Due to their disruptive nature, predicting cyber attacks is an important research effort. Most research efforts focus on using network traffic for forecasting as in\cite{park2012cyber, pontes2011applying}. These methods leverage network traffic or sensors at different layers as the underlying data to forecasting models. We specifically avoid network data and base our predictions on open source information. Other efforts include \cite{zhang2015predicting} which only used the National Vulnerability Database with moderate success and they highlight the difficulty in using public sources for building effective models. The main difference is that our work is based on actual cyber event ground truth as reported by the two target organizations. The closest to our research is Gandotra et al \cite{comptech} who outlined a number of cyber prediction efforts using statistical modeling and algorithmic modeling. They highlight several significant challenges that we tried to address. The first challenge is that open source ground truth is often incomplete and should be compiled from multiple sources and analysis doesn't scale to real world scenarios.  We were able to get ground truth data from two companies, this ground truth is across three different attack vectors and is over a two year time period. The additional challenges in \cite{comptech} focus on the volume, speed and heterogeneity of network data which we avoid since we are attempting to prevent cyber events specifically with non-network data. They also present two modeling approaches of statistical modeling and algorithmic modeling. We used statistical models not unlike what they present as classical time series models with auto-regressive, integrated moving average with historical data and external signals.

Developing a precise model for the dynamic behavior of time series is a challenging problem and an essential one for the success of forecasting methods. Researchers have extensively studied and used time-series analysis in many domains, such as finance~\cite{lendasse2000non}, epidemiology~\cite{chakraborty2014forecasting,Wang:2015:DPA}, geophysics~\cite{shumway2010time}, and sociology~\cite{box2014time}. A popular strategy for analyzing time series data is using  classical autoregressive models such as AR, ARMA, ARIMA, and ARIMAX~\cite{box2015time,shumway2010time,prado2010time}. Autoregressive models are widely used in intrusion detection, detecting DoS attacks, and network monitoring~\cite{viinikka2009processing}. These models assume that the underlying data-generating process is linear, \textit{i.e.}, the value at a time point is a linear combination of the past values. However, real-world time series exhibit volatility and nonlinearity. A way to deal with the problem of volatility is to employ ARCH and GARCH, which are extensions of classical autoregressive models~\cite{douc2014nonlinear}. 

Neural network based models have been widely used for time series prediction tasks such as weather forecasting~\cite{xingjian2015convolutional}, sentence completion~\cite{mirowski2015dependency} and oilfield production prediction~\cite{cheung2017oreonet}. Recent success in recurrent neural networks~\cite{medsker2001recurrent} on such tasks has led to produce several variants of the models such as Long Short Term Memory~\cite{hochreiter1997long}, Peephole-LSTM~\cite{gers2002learning}, Depth Gated RNN~\cite{yao2015depth}, Clockwork RNN~\cite{koutnik2014clockwork}, Gated Recurrent Unit (GRU)~\cite{cho2014learning} and Phased LSTM~\cite{neil2016phased}. Cyber security community recently adapted neural network models. DeepExploit~\cite{tavabi2018dark} uses neural embeddings to predict exploit likelihood of a vulnerability. Filonov \emph{et al}~\cite{filonov2017rnn} developed RNN based model for early detection of cyber attacks. Our model, on the other hand, considers various recurrent neural network based models on external signals to identify predictive signals and we report results on GRUs only. 
\section{Conclusions}
In this paper, we tackle the challenging problem of predicting targeted cyber attacks. Cyber attacks do not emerge randomly, rather, they are caused by a vast amount of hidden factors which include motivational factors like financial, espionage fun or grudge; the exploitation of new or known software and hardware vulnerabilities; the appearance of new vulnerabilities and malwares, to list a few. 
Our approach is to systematically, and in a fully automated manner, harness these hidden factors to identify predictive signals from various public data sources such as social media signals, Internet-based sensors, dark Web, blogs, and more. 

We used state of the art machine learning methods for time series prediction. We showed that  historical ground truth data already carried enough information to enable these methods to learn patterns to predict future events. We then showed how incorporating information from external signals improves on these forecasts and quantified the improvement. In this manner, we are able to identify the best signals to predict cyber threats for each target organization. 

Our framework provides a systematic way that can be used to improve decision making in Cyber Security Policy. Our results show that depending on the specific target and type of attack, different data sources should be monitored for the early prediction and mitigation of cyber threats. Indeed, while some external signals are good predictors for both organizations, the best performing signals are unique to each target. Thus, providing suggestive evidence that our method is able to recognize the idiosyncrasies and specific vulnerabilities of each organization.  

Future work will be devoted to enhancing the predictive power of external signals. One direction is by inferring latent factors that are common to all signals that have better predictive power than the signal by itself. A second idea is to evaluate various linear combinations of predictive signals as well as exploiting the semantics of the identified signals. Signal fusion may be especially useful for weekly-level predictions, as different signals may complement each other and allow for more robust performance over time. 



\appendix

Supplemental material can be found here: \url{http://bit.ly/2GZWFdZ}

\IEEEdisplaynontitleabstractindextext

%
\IEEEpeerreviewmaketitle

\ifCLASSOPTIONcompsoc
  \section*{Acknowledgments}
\else
  \section*{Acknowledgment}
\fi

{\footnotesize This work was supported by the Office of the Director of National Intelligence (ODNI) and the Intelligence Advanced Research Projects Activity (IARPA) via the Air Force Research Laboratory (AFRL) contract number FA8750-16-C-
0112. The U.S. Government is authorized to reproduce and distribute reprints for Governmental purposes notwithstanding any copyright annotation thereon. Disclaimer: The views and conclusions contained herein are those of the authors and should not be interpreted as necessarily representing the official policies or endorsements, either expressed or implied, of ODNI, IARPA, AFRL, or the U.S. Government.
}

\ifCLASSOPTIONcaptionsoff
  \newpage
\fi

\balance




\bibliographystyle{IEEEtran}
\bibliography{dnn}

\end{document}